\documentstyle[12pt]{article}

\newcommand{\vmax}{v_{\rm max}}
\newcommand{\rmin}{r_{\rm min}}

\newcommand{\vs}{v_{\rm s}}

\newcommand{\eV}{{\rm eV}}
\newcommand{\sech}{{\rm sech}}
\newcommand{\disp}{\displaystyle}

\newcommand{\bibn}[1]{\frac{\rm d}{{\rm d} #1}}
\newcommand{\bibnB}[2]{\frac{{\rm d} #1}{{\rm d} #2}}
\newcommand{\bibnNkaiB}[3]{\frac{{\rm d}^#3 #1}{{\rm d} #2^#3}}
\newcommand{\henb}[1]{\frac{\partial}{\partial #1}}
\newcommand{\henbB}[2]{\frac{\partial #1}{\partial #2}}
\newcommand{\henbNkaiB}[3]{\frac{\partial^#3 #1}{\partial #2^#3}}

\begin{document}

\begin{center}
{\large \bf Semi-Phenomenological Analysis of Dynamics\\
of Nonlinear Excitations\\
in One-Dimensional Electron-Phonon System }

\vspace{1cm}

Makoto {\sc Kuwabara}
\footnote{Present address: Fundamental Physics Section,
Physical Science Division,
Electrotechnical Laboratory, 
Umezono 1-1-4, Tsukuba, Ibaraki 305, Japan}, 
Akira {\sc Terai}$^*$ and 
Yoshiyuki {\sc Ono} 

\vspace{1cm}

{\sl 
Department of Physics, Toho University, \\
Miyama 2 - 2 - 1, Funabashi, Chiba 274, Japan\\
$^*$Department of Physics, Kitasato University\\
Kitasato 1-15-1, Sagamihara, Kanagawa 228, Japan
}

\end{center}

\vspace{1cm}

{\bf abstract}\\
The structure of moving nonlinear excitations in one-dimensional 
electron-phonon systems is studied semi-phenomenologically by using 
an effective action in which the width of the nonlinear excitation is treated 
as a dynamical variable.  
The effective action can be derived from Su, Schrieffer and Heeger's 
model or its 
continuum version proposed by Takayama, Lin-Liu and Maki with an assumption 
that the nonlinear excitation moves uniformly without any deformation 
except the change of its width.  
The form of the action is essentially the same as that discussed by Bishop and 
coworkers in studying the dynamics of the soliton in polyacetylene, 
though some details are different.  
For the moving excitation with a velocity $v$, the width is determined by 
minimizing the effective action.  
A requirement that there must be a minimum in the action as a function of its 
width provides a maximum velocity.  
The velocity dependence of the width and energy can be determined.  
The motions of a soliton in polyacetylene and an acoustic polaron in 
polydiacetylene are studied within this formulation.  
The obtained results are in good agreement with those of 
numerical simulations.

\newpage
\section{Introduction}
It is generally accepted that dynamical behaviors of nonlinear excitations 
in one-dimensional electron-phonon systems such as conjugated polymers 
play an important role in determining physical properties of those 
systems.~\cite{Polymer_Springer}  
The motions of the nonlinear excitations have been studied by analytic 
methods~\cite{Geraldo,Wilson} 
and by numerical simulations.~\cite{SS,Mele,Bishop,Terai}  
In recent papers,~\cite{SdynI,SdynII,SdynIII,SdynIV,amp} 
using Su, Schrieffer and Heeger's (SSH) model~\cite{SSH} modified to 
include an external electric field, we have investigated the dynamics of a 
charged soliton in polyacetylene by numerical simulations in the presence of 
an electric field.  
We have found the following results: 
i) The soliton has a saturation velocity of about four times the sound 
velocity.~\cite{Bishop,SdynI}  
ii)~The width of the soliton decreases as the soliton velocity 
increases.~\cite{SdynII}  
Furthermore, the time dependence of the width determined from the spatial 
variation of the bond order parameter involves an oscillatory component, 
which is related to the amplitude mode around the 
soliton.~\cite{amp}  
iii) The overall behavior of the total energy of the system as a function 
of the soliton velocity $v$ is well expressed as 
\begin{equation}
\Delta \varepsilon_{\rm tot}(v)
 = -(M_{\rm s}v_{\rm m}^2 / 2) \ln (1-v^2/v_{\rm m}^2)
\label{eq:Et_num}
\end{equation}
with $\Delta \varepsilon_{\rm tot}(v) = 
\varepsilon_{\rm tot}(v)-\varepsilon_{\rm tot}(0)$, 
where $v_{\rm m}$ is the maximum velocity and $M_{\rm s}$ the soliton 
effective mass of about four times the bare electron mass.~\cite{SdynIII}  

Bishop and coworkers~\cite{Bishop} also have studied the dynamics of the 
soliton with numerical simulations and shown the 
saturation of the soliton velocity whose value is in rough agreement with our 
numerical result.  
Furthermore they have given a phenomenological explanation for the saturation 
mechanism 
that the time necessary for an ion group on a lattice point to respond to 
the reverse of the 
dimerization pattern caused by the soliton translation can not be shorter than 
the characteristic time of the lattice vibrations, $\omega_0^{-1}$.  
Here $\omega_0$ is the renormalized optical phonon frequency.  
Alternatively they have explained the saturation by another independent 
method.  
According to that method, the width $\xi$ of the soliton moving with a 
velocity $v$ is determined by minimizing an effective action.  
A requirement that the action must have an minimum as a function of $\xi$ 
provides the saturation (maximum) velocity.  

In this paper, we reconsider the phenomenological method for describing the 
structure of 
moving nonlinear excitations in one-dimensional electron-phonon systems by 
using the effective action originally proposed by Bishop and coworkers.  
This method is applicable not only to the soliton in polyacetylene but also 
to a polaron or other nonlinear excitations in conjugated polymers.  
The effective action is made up of a potential energy part and a kinetic one, 
and the width of the nonlinear excitation is regarded as a dynamical 
variable.  
The potential energy is defined by the energy of a static nonlinear 
excitation for each value of the width.  
Both parts are calculated numerically or analytically by using the SSH 
model~\cite{SSH} or its continuum version, {\it e.g.} Takayama, Lin-Liu, 
and Maki's (TLM) model in the case of the soliton in 
polyacetylene.~\cite{TLM}  
Using the effective action, we determine the saturation velocity, the relation 
between the velocity and the width, and the velocity dependence of the energy 
of the nonlinear excitations.  
We concentrate on the dynamics of a soliton in polyacetylene and of an 
acoustic polaron in polydiacetylene.~\cite{Cade}  
The latter is formed by interactions between a free electron and acoustic 
phonons, and its dynamical behavior has been investigated analytically by 
Wilson.~\cite{Wilson}  
The width and the energy of the acoustic polaron have been presented as the 
function of its velocity.  
For both cases, we show that the dynamical behaviors evaluated from the action 
are in good agreement with those obtained from the numerical simulations.  
Furthermore, for the case of the acoustic polaron, it is indicated that the 
results derived from the effective action are consistent with the analytic 
results by Wilson.  

This paper is organized as follows.  
In $\S$ 2 we define the effective action for the moving soliton and 
its structure is analyzed.  
The results are compared with those obtained by the numerical simulations.  
The same analyses are performed for the acoustic polaron in $\S$ 3.  
We present conclusions and discuss the results in the final section.  

\section{Structure of Moving Soliton in Peierls-Dimerized Linear Chain}
\subsection{Effective action}
In order to discuss the effective action describing the behavior of a moving 
soliton, we start with the SSH model,  
\begin{equation}
H_{\rm SSH} = -{\sum_{n\sigma}}(t_0-\alpha y_n)[c^\dagger_{n\sigma}
                                            c_{n+1\sigma}+ h.c. ] 
            +\frac{K}{2}{\sum_n}y_n^2
            +\frac{M}{2}{\sum_n}\dot{u}_n^2,
\end{equation}
where $y_{n}=u_{n+1}-u_{n}$.  
The quantity $t_0$ is the nearest neighbor transfer integral of 
$\pi$-electrons in the regular lattice, $\alpha$ the electron-lattice coupling 
constant, and $u_n$ the lattice displacement of the $n$-th site from its 
equidistant position.  
The operators $c^{\dagger}_{n\sigma}$ and $c_{n\sigma}$ create and 
annihilate a $\pi$-electron with spin $\sigma$ at the $n$-th site, 
respectively.  
$K$ is the force constant mainly due to the $\sigma$-bond, $M$ the mass of a 
CH unit.  

We consider a uniformly moving soliton with a velocity $v$.  
The soliton is regarded as a spatially extended object with an internal 
degree of freedom characterized by its width.  
The lattice displacement at the $n$-th site is assumed to be expressed as,
\begin{equation}
u_n = (-)^n u_0 \tanh \left( \frac{na-vt}{\xi} \right),
\label{eq:u_n}
\end{equation}
where $u_0$ denotes the amplitude of the staggered lattice displacement in the 
perfectly dimerized state, $a$ the lattice constant, and $\xi$ the soliton 
width.  
Here, in a collective-coordinate theory, the center of mass of the soliton and 
its width are treated as dynamical variables characterizing the structure of 
the moving soliton.~\cite{Rice-Mele,Rice,Rice-Jeyadev-Phillpot}  
These two collective coordinates depend on time.  
In the present formulation the center of the soliton moves with the velocity 
$v$ and the width is assumed to be determined by the velocity.  
We ignore any other internal excitations such as the amplitude mode.  

The effective action is defined by the potential energy minus the kinetic 
energy,
\begin{equation}
A(r;v) = V(r) - \frac{\beta_v}{r},
\label{eq:A_sol}
\end{equation}
with
\begin{equation}
r = \frac{\xi(v)}{\xi_0},
\mbox{\hspace{1cm}}
\beta_v = \frac{2aKu_0^2}{3\xi_0} 
         \left( \frac{v}{v_{\rm s}} \right)^2,
\label{eq:beta_v}
\end{equation}
where $\xi_0$ is the width of a stable static soliton and $v_{\rm s}$ the 
sound velocity ($=a\sqrt{K/M}$).  
$V(r)$ denotes a potential energy which is 
given by the sum of the electronic energy and the lattice elastic energy of 
the system containing a soliton with the relative width $r$.  
The second term on the r.h.s. of eq. (\ref{eq:A_sol}) 
comes from the lattice kinetic energy.  
The details of the derivation of the effective action are given in 
Appendix.  
In Fig. 1 the potential energy $V(r)$ is shown, where the optimized energy 
at $r=1$ is taken to be zero.  
We can neglect the constant part independent of $r$ because in the present 
method it is essential to find a minimum of the action as a function of $r$.  

The action can also be defined from the continuum version of the SSH model, 
i.e. the TLM model.  
In this case $V(r)$ is essentially identical with the creation energy of a 
static soliton for the fixed relative width $r$, and the coefficient of the 
kinetic energy term is expressed as 
\begin{equation}
\beta_v = \frac{2v^2\Delta_0^3}{3\pi\lambda v_{\rm F}^2 \omega_{\rm Q}^2},
\label{eq:beta_v_cont}
\end{equation}
where $2\Delta_0$ is the Peierls gap, $\lambda (=\alpha^2/Kt_0)$ 
the dimensionless 
electron-phonon coupling constant, $v_{\rm F}$ the Fermi velocity and 
$\omega_{\rm Q} (=\sqrt{4K/M})$ the bare optical phonon frequency.  
It is straightforward to confirm that $\beta_v$ in eq. (\ref{eq:beta_v}) 
reduces to that in eq. (\ref{eq:beta_v_cont}) in the continuum limit 
( or in other words, the weak coupling limit).  

Here we note that, in the study by Bishop and coworkers,~\cite{Bishop} 
$V(r)$ is approximated by a quadratic function of $r$ which coincides to 
$V(r)$ calculated from the TLM model at two points $r=0$ and 1.
Its explicit $r$-dependence is shown in Fig. 1 by a dashed curve.  
Then the effective action is rewritten as,~\cite{footnote1}
\begin{equation}
A = \frac{2}{\pi}\Delta_0 + \Delta_0 (1-\frac{2}{\pi})(1-r)^2 
                    -\frac{\beta_v}{r} ,
\label{eq:A_sol2}
\end{equation}
The first term on the r.h.s. is the energy of the soliton at rest.  
However this approximation is too crude for our purposes as will be seen 
in the following.  

\subsection{Saturation of velocity}
The maximum velocity is determined from the requirement that the action 
must have at least one minimum as a function of $r$.  
If the approximation eq. (\ref{eq:A_sol2}) is used, the requirement is 
fulfilled when $\beta_v \leq \beta_{\rm c} \equiv (2/3)^3(1-(2/\pi))$.  
For $\beta_v=\beta_{\rm c}$, the minimum point of the action is located at 
$r=2/3$.  
From $\beta_{\rm c}$, the maximum velocity is estimated to be
\begin{equation}
\vmax = \frac{2}{3} \sqrt{\pi\lambda (1-\frac{2}{\pi})} 
                           v_{\rm F} \omega_{\rm Q} / \Delta_0 
          \simeq 3.74 v_{\rm s}
\end{equation}
This velocity is appreciably larger than the saturation velocity 
($\simeq 3.22v_{\rm s}$) obtained from the numerical simulation in the 
previous works.~\cite{SdynIII}  

Next we use the exact expression of $V(r)$ defined from the SSH model 
in order to calculate $\beta_{\rm c}$, the critical value of $\beta_v$.  
$\beta_{\rm c}$ is defined as a value where the minimum of the derivative of 
the action with respect to  $r$ becomes zero, and is calculated numerically.  
In Fig. 2 typical examples of the action and its derivative are shown for 
$\beta_v < \beta_{\rm c}$ (a), $\beta_v = \beta_{\rm c}$ (b) and 
$\beta_v > \beta_{\rm c}$ (c).  
From $\beta_{\rm c}$, the maximum velocity is evaluated to be 
$\vmax = 3.18v_{\rm s}$, which is in good agreement with that obtained 
in the numerical simulation.  
When the action defined from the TLM model is used we obtain also a rather 
good result, $\vmax = 3.22v_{\rm s}$.  

\subsection{Velocity dependence of the soliton width}
The width of the moving soliton is determined from the 
condition of minimizing the effective action, as described above.  
The velocity dependence of the width obtained by using the effective action 
defined from the SSH model is shown in Fig. 3 along with the result of the 
numerical simulation for comparison.  
For the numerical result we used the width calculated from the bond order 
parameter.~\cite{SdynII}  

The relation between the width and the velocity obtained from the action is in 
good agreement with that of the numerical simulation.  
We fit the relations to the following form through the least square fitting 
in the region $v < 2\vs$,
\begin{equation}
r = 1 -a \left( \frac{v}{v_{\rm s}} \right)^2
      -b \left( \frac{v}{v_{\rm s}} \right)^4,    \label{eq:r-v}
\end{equation}
and obtain for the analytic curve, 
\begin{eqnarray}
a_{\rm analy} & = & 2.4 \times 10^{-2}, \nonumber \\
b_{\rm analy} & = & 5.3 \times 10^{-4}, \nonumber 
\end{eqnarray}
and for the curve of the numerical simulation,
\begin{eqnarray}
a_{\rm num} & = & 1.6 \times 10^{-2}, \nonumber \\
b_{\rm num} & = & 9.7 \times 10^{-4}. \nonumber 
\end{eqnarray}

The small difference between the analytic and numerical results may be 
due to the inaccuracy in the velocity and the width calculated from the 
lattice displacement pattern in the numerical simulation.  
These quantities have a rather 
poor accuracy 
due to the tail structures in the lattice displacement pattern after the 
soliton has traveled a rather long distance.  
The tail structures may particularly give an inaccuracy to the width.  
Taking account of this inaccuracy, the agreement between the 
semi-phenomenological analysis and the numerical simulation seems to be 
satisfactory.  

\subsection{Velocity dependence of energy}
The energy of a moving soliton is represented by
\begin{equation}
E(v) = V(r) + \frac{\beta_v}{r},
\label{eq:Et_anly}
\end{equation}
where the relative width $r$ is the function of the velocity $v$ as shown 
in Fig. 3.  
Figure 4 indicates the velocity dependence of the soliton energy, in which 
results of the numerical simulation are also shown for the total and the 
lattice kinetic energies of the system containing the moving soliton.  
Although a slight difference in magnitude between the analytic result for 
the total energy and that of the simulation is seen, the functional form of 
the analytic result reproduces well the numerical one.  
Here the energies calculated in the simulation involve the contributions of 
the 
lattice vibrations (tail structure) emitted from the moving soliton; this 
contribution is not directly related to the energy of the soliton.  
If this effect is taken into account, the slight difference would become 
smaller.  

In our previous paper, we have shown that the energy of the moving soliton 
looks to diverge logarithmically as a function of the velocity 
(eq. (\ref{eq:Et_num})).  
Here we deal with the behavior of the energy around the maximum velocity, 
using eq. (\ref{eq:Et_anly}).  
The derivative of $E(v)$ with respect to $v$ is expressed as
\begin{eqnarray}
\bibnB{E(v)}{v} & = & \bibnB{r}{v} 
   \left( \bibnB{V}{r} - \frac{\beta}{r^2} v^2 \right) + \frac{2\beta}{r} v,
                                        \nonumber \\
                & = & -\frac{2\beta}{r^2} v^2 \bibnB{r}{v} 
                      + \frac{2\beta}{r} v,  \label{eq:dE/dv}
\end{eqnarray}
where $\beta$ is a constant defined by $\beta =\beta_v /v^2$.  
In deriving the last expression we have used the relation between $r$ and $v$,
\begin{equation}
\henbB{A}{r} = \bibnB{V}{r} + \frac{\beta}{r^2} v^2 = 0. 
\label{eq:dA/dr}
\end{equation}
The velocity derivative of $r$ is obtained by differentiating eq. 
(\ref{eq:dA/dr}) by $v$,
\begin{equation}
\bibn{v} \left( \henbB{A}{r} \right) = 
\bibnB{r}{v} \left( \bibnNkaiB{V}{r}{2} - \frac{2\beta}{r^3}v^2 \right)
 + \frac{2\beta}{r^2}v = 0.
\label{eq:d2A/dvdr}
\end{equation}
On the other hand, when $v=\vmax$ ($\beta_v=\beta_{\rm c}$), it is 
obvious from Fig. 2(b) that $\disp\henbB{A}{r}$ and 
$\disp\henbNkaiB{A}{r}{2}$ satisfy following relations,
\begin{eqnarray}
\left. \henbB{A}{r} \right|_{r=r_{\rm min}} & = & 0,
\label{eq:dA/dr(rmin)}
\\
\left. \henbNkaiB{A}{r}{2} \right|_{r=r_{\rm min}}
&=&\bibnNkaiB{V(r_{\rm min})}{r}{2} - \frac{2\beta}{r_{\rm min}^3}v_{\rm max}^2
= 0.
\label{eq:d2A/dr2}
\end{eqnarray}
From eqs. (\ref{eq:d2A/dvdr}) and (\ref{eq:d2A/dr2}) the derivative of $r$ 
with respect to $v$ diverges to minus infinity at $v=\vmax$.  
Then $\disp\bibnB{E}{v}$ diverges as $v$ approaches $\vmax$.  Noting that 
the value of $r$ itself does not diverge at $v=\vmax$, we can safely assume 
that $r(v)=r(\vmax)+C_0(v-\vmax)^\mu$ with $0 < \mu <1$ near $\vmax$. 
This assumption will be confirmed by the result. 
If we put $r(v)=r(\vmax)+\Delta r$, then $\Delta r$ can be estimated 
from eq.~(\ref{eq:dA/dr}) in the following way. 

We expand the expression for $\disp\henbB{A}{r}$ 
in terms of $\Delta r$ and $\Delta v (=v-\vmax)$ up to second order. 
\begin{eqnarray}
\left. \henbB{A}{r} \right|_{\rmin(v)} & \simeq &
\frac{1}{2} \left. \henbNkaiB{A}{r}{3} \right|_{\rmin(\vmax)}
 (\Delta r)^2 
+\frac{2\beta\vmax}{[\rmin(\vmax)]^2} \Delta v
\nonumber \\
& & +\frac{\beta}{\rmin^2} (\Delta v)^2 
    -\frac{4\beta\vmax}{\rmin^3} (\Delta r)(\Delta v)
\nonumber \\
& = & 0,
\end{eqnarray}
where we have used eqs. (\ref{eq:dA/dr(rmin)}) and (\ref{eq:d2A/dr2}). 
It is obvious from what we have assumed for $\Delta r$ that the last 
two terms in the above expansion are irrelevant. 
It is also clear from the general behavior of $A(r)$ that 
\begin{equation}
\left. \henbNkaiB{A}{r}{3} \right|_{\rmin(\vmax)} > 0.
\end{equation}
As a result, we end up with 
\begin{equation}
\Delta r \propto \sqrt{\vmax -v},
\end{equation}
consistently with the first assumption. 
Therefore we can conclude that
\begin{equation}
E(v) \simeq E(\vmax) - C \sqrt{\vmax-v},
\end{equation}
by combining the above result for $\Delta r$ with eq.~(\ref{eq:dE/dv}). 
$E(\vmax)$ itself is finite but $\disp\bibnB{E}{v}$ diverges at $v=\vmax$.  
This singularity would have been seen as a logarithmic singularity when we 
analyzed the overall energy-velocity relation obtained from the numerical 
simulation.  

\section{Dynamics of Acoustic Polaron}
\subsection{Effective action}
An acoustic polaron can be described in terms of the SSH Hamiltonian by 
postulating that the system has only one electron.  
In Fig. 5 the self-consistent solution is shown for the charge distribution, 
the lattice displacement and the bond order parameter, respectively.  
The values of the parameters used are $t_0=0.64\eV$, 
$\alpha=0.38\eV/{\rm \AA}$, $K=2.4\eV/{\rm \AA}^2$, $a=4.9{\rm \AA}$, 
$\omega_{\rm Q}=0.96\times 10^{-2}\eV$, which might be appropriate for 
polydiacetylene.  
The periodic boundary condition is assumed.  
In an infinite system, a static acoustic polaron can be well 
expressed by the form,~\cite{Wilson} 
\begin{equation}
u_n = -\frac{u_0}{r} \tanh \left( \frac{na}{\xi_0 r} \right),
\label{eq:u_pl_st}
\end{equation}
with $\xi_0\ (=\disp\frac{\hbar^2}{m} \frac{M\vs^2}{2\alpha^2 a^3})$ the width 
of a stable acoustic polaron and 
$u_0\ (=2\alpha a^2/M\vs^2)$ the amplitude of the lattice displacement at a 
long distance from the center in the case of a stable polaron,~\cite{Wilson} 
where we assume the center to be at the origin.  
For a stable polaron the relative width $r$ is equal to unit.  
The prefactor in eq. (\ref{eq:u_pl_st}) is determined so as to minimize the 
total energy of the system when the polaron width $\xi_0 r$ is fixed.  
The change of the lattice spacing affects directly the transfer integral of 
the electron,
\begin{equation}
y_n = u_{n+1} - u_n
    \simeq -\frac{au_0}{\xi_0 r^2} \sech^2 \left( \frac{na}{\xi_0 r} \right).
\label{eq:y_pl_st}
\end{equation}
Figure 6 indicates the acoustic polaron in the infinite system expressed 
by $u_n$ and $y_n$.  
In the numerical simulations of the motion of the acoustic polaron, we use a 
finite size system and impose the periodic boundary condition.  
This condition leads to additional terms in eqs. (\ref{eq:u_pl_st}) and 
(\ref{eq:y_pl_st}).
\begin{eqnarray}
u_n & = & -\frac{u_0}{r} \tanh \left( \frac{na}{\xi_0r} \right) + C n, \\ 
y_n & = & -\frac{au_0}{\xi_0r^2} \sech^2 \left( \frac{na}{\xi_0r} \right) + C,
\label{eq:y_pl_st_pbc}
\end{eqnarray}
with
\begin{equation}
C = \frac{au_0}{N\xi_0r^2} \sum_{n=-N/2}^{N/2} 
       \sech^2 \left( \frac{na}{\xi_0r} \right)
  \simeq \frac{2u_0}{Nr}.
\end{equation}

A uniformly moving acoustic polaron will be described in the $u_n$ 
representation as 
\begin{equation}
u_n = -\frac{u_0}{r} \tanh \left( \frac{na-vt}{\xi_0r} \right) 
      +\frac{C}{a} (na-vt).
\end{equation}
Inserting this expression into the kinetic energy part in the SSH model, we get
\begin{eqnarray}
E_{\rm K} = \frac{M}{2} \sum_n \dot{u}_n^2 & \simeq & 
 \frac{M}{2a} v^2 u_0^2 \int_{-Na/2}^{Na/2} dx 
     \left\{ \frac{1}{\xi_0r^2} \sech^2 \left( \frac{x-vt}{\xi_0r} \right) 
             -\frac{2}{Nar}   \right\}^2. \nonumber  \\
 & = & \frac{2Mv^2u_0^2}{a} \left( \frac{1}{3\xi_0r^3} -\frac{1}{Nar^2} \right)
\label{eq:EK_Pol}
\end{eqnarray}
The effective action for the moving acoustic polaron is defined similarly to 
the case of the soliton as,
\begin{equation}
A = V(r) -\frac{2Mv^2u_0^2}{a}  
                \left( \frac{1}{3\xi_0r^3} - \frac{1}{Nar^2} \right).
\end{equation}
The potential term $V(r)$ is calculated by substituting 
eq. (\ref{eq:y_pl_st_pbc}) into the SSH model in the same way as for the 
soliton in polyacetylene.  
We show the potential $V(r)$ in Fig. 7.  

\subsection{Saturation of velocity}
The saturation of the velocity in the case of the acoustic polaron has been 
discussed by Wilson and estimated to be nearly equal to the sound 
velocity.~\cite{Wilson}  
The dynamical simulation of the acoustic polaron in the presence of an 
electric field shows that it is about $0.8\sim 0.9v_{\rm s}$.~\cite{Arikabe}  
By using the effective action it is found that the acoustic polaron becomes 
zero width at the sound velocity as is shown in the next section.  
Then the maximum velocity is expected to be the sound velocity from this 
fact.  
The reason why it is slightly smaller in the numerical simulation will be that 
the computing time is finite and also that the lattice vibrations excited by 
the motion of the polaron may be acting as scatterers yielding a small 
resistance against the motion.  

\subsection{Velocity dependence of width}
The solid line in Fig. 8 indicates the relation between the width and the 
velocity of the acoustic polaron determined from the action.  
The data of the numerical simulation is also 
shown.~\cite{Arikabe}  
The agreement between the analytic result~\cite{footnote3} 
and the numerical one is quite 
good.  
It seems that the width of the acoustic polaron becomes zero at 
$v=v_{\rm s}$.  
According to the analytical study by Wilson,~\cite{Wilson} who used instead of 
the effective action the equation of motion for the lattice displacement and 
the Schr\"odinger equation for the electronic wave function in the continuum 
limit, the velocity dependence of the width is of the form
\begin{equation}
r = 1 - \left( \frac{v}{v_{\rm s}} \right)^2,
\label{eq:wid-v;Wilson}
\end{equation}
which is indicated by the broken line in Fig. 8.  
Furthermore it has been shown that the electronic energy and the lattice 
potential energy as well as the lattice kinetic energy are expressed as the 
functions of $r$ as\cite{Wilson}
\begin{eqnarray}
\epsilon_{\rm e} & = & -\frac{U}{r^2}, \\
\epsilon_{\rm LP} & = & \frac{2}{3} \frac{U}{r^3}, \\
\epsilon_{\rm LK} & = & \frac{2}{3} \frac{U}{r^3} 
                        \left( \frac{v}{v_s} \right)^2 .  \label{eq:EK_Pol2}
\end{eqnarray}
Here 
\begin{equation}
U = \frac{m}{\hbar^2} \frac{8\alpha^4a^6}{M^2v_{\rm s}^4}.
\end{equation}
If we use these expressions, we obtain the same result as 
eq. (\ref{eq:wid-v;Wilson}) with the help of the effective action, 
which supports the appropriateness of the semi-phenomenological method.  
In fact it is straight forward to confirm that 
the lattice kinetic energy eq.(\ref{eq:EK_Pol}) reduces to 
eq.(\ref{eq:EK_Pol2}) in the infinite system.  

Here note that we recognize an appreciable difference near the sound 
velocity between the velocity dependences of the widths calculated 
from the action and from the continuum approximation by Wilson in Fig. 8.  
This may be due to the discreteness of the SSH model, 
since the width of the acoustic polaron becomes quite narrow near 
the sound velocity.  

\subsection{Velocity dependence of energy}
The energy of the moving acoustic polaron is given by
\begin{equation}
E(v) = V(r) + E_{\rm K}.
\end{equation}
The velocity dependence of the energy $E(v)$ and that of the kinetic 
energy part $E_{\rm K}$ are shown in Fig. 9 together with the results of the 
numerical simulations.  
Both of the energies diverge as the velocity approaches the sound velocity.  
The agreements between the analytic and numerical results are excellent.  
These results agree also with those obtained by Wilson.  
This fact along with the results in the previous subsections indicates that 
the present method in terms of the effective action is essentially equivalent 
to Wilson's analytic method and that both methods can reproduce the results 
of the numerical simulation quite well.  

\section{Summary and Discussion}
We studied the structure of a moving soliton and a moving acoustic polaron in 
the one-dimensional electron-phonon system by means of a semi-phenomenological 
method by introducing the effective action.  
The effective action for a moving nonlinear excitation consists of the 
potential energy with its width as the ``coordinate'' and of the kinetic 
energy.  
In the continuum limit (the TLM model) for the soliton problem, 
the potential energy is expressed by 
the creation energy of the soliton with a given width.  
It is assumed that the width can be regarded as a dynamical variable which 
depends on the velocity.  
The widths of the moving nonlinear excitations are determined so as to 
minimize the effective action.  
The requirement that the action must have a minimum gives the upper limit of 
the velocity, and its value is found to be nearly the same as 
the saturation velocity obtained by the numerical simulation.  
The velocity dependence of the width and that of the energy calculated by the 
semi-phenomenological method present almost the same behaviors as the results 
obtained from the dynamical simulations.  
We may reasonably conclude that the dynamical behaviors of the soliton and the 
acoustic polaron can be described by the phenomenological effective action.  

From the analysis using the effective action, it is found that for the case of 
the soliton there occurs the divergence not in the energy itself but in its
derivative with respect to the velocity at $v=v_{\rm max}$.  
In contrast to the case of the soliton, the energy of the acoustic polaron 
diverges at $v=v_{\rm s}$.  

The dynamical behavior of the acoustic polaron is described quite well by the 
effective action while such a description was not much accurate in the case of 
the soliton in polyacetylene.  
One of the reasons for this is the acoustic polaron has no extra internal 
excitation 
in contrast to the case of the soliton which has several localized modes 
omitted in the present treatment.  

In this paper we have focused our attention on the soliton and the 
acoustic polaron.  
The semi-phenomenological method developed in this work is applicable to the 
cases of other nonlinear excitations in one-dimensional electron-phonon 
systems such as an optical polaron.  
We are now studying this subject, the result of which will be presented 
elsewhere.  

\section*{Acknowledgements}
The authors appreciate fruitful discussions with Professor T. Ohtsuki.  
They also thank Y. Arikabe and Y. Hori for useful discussions.  
This work was partly financed by Grant-in-Aid for 
Scientific Research from the Ministry of Education, 
Science and Culture, No. 05640446.

\vspace{1cm}
\noindent
{\large \bf Appendix: Derivation of the Effective Action}\\
We show the derivation of the effective action for the soliton.  
Although we treated the soliton width as a function of the velocity 
in the previous section, we consider it as a time depending variable here.  

Using eq. (\ref{eq:u_n}), the lattice kinetic energy is 
\begin{equation}
E_{\rm K} = \frac{M}{2} \sum \dot{u}_n^2 
  = \frac{Mu_0^2}{2a} \left( \frac{4}{3} \frac{v^2}{\xi}
                            +\frac{\pi^2-6}{9} \frac{\dot{\xi}^2}{\xi} \right).
\end{equation}
The Lagrangian for the fixed velocity $v$ is expressed as
\begin{equation}
L(\xi,\dot{\xi}) = \frac{M_{\rm s}(\xi)}{2} v^2
                 +\frac{M_{\rm i}(\xi)}{2} \dot{\xi}^2 - V(\xi),
\label{eq:L}
\end{equation}
with
\begin{equation}
M_{\rm s}(\xi) = \frac{4}{3} \frac{Mu_0^2}{a\xi},
\mbox{\hspace{1cm}}
M_{\rm i}(\xi) = \frac{\pi^2-6}{9} \frac{Mu_0^2}{a\xi},
\end{equation}
where $V(\xi)$ denotes the potential energy 
which is given by the sum of the electronic and lattice elastic energies.  
Then the equation of motion is written as
\begin{equation}
M_{\rm i}(\xi)\ddot{\xi} = \frac{1}{2} \henbB{M_{\rm s}(\xi)}{\xi} v^2
  -\frac{1}{2} \henbB{M_{\rm i}(\xi)}{\xi} \dot{\xi}^2
  -\henbB{V(\xi)}{\xi}.
\label{eq:eq_mtn_wid}
\end{equation}
There can be the amplitude mode around the soliton, which leads to an 
oscillation 
of the soliton width.  
Therefore, the equation of motion must have an oscillatory solution.  
If the equation of motion eq. (\ref{eq:eq_mtn_wid}) 
is solved, the amplitude mode 
frequency may be evaluated as a function of the velocity, which we have 
studied numerically in the previous paper.~\cite{amp}  
We have confirmed that eq. (\ref{eq:eq_mtn_wid}) yields a reasonable value 
for the frequency of the amplitude mode in the case of the static soliton 
($v=0$).~\cite{Terai-Ono}  
In the case of the moving soliton ($v\neq 0$), however, the amplitude mode 
frequency derived from eq. (\ref{eq:eq_mtn_wid}) decreases with the velocity 
in contrast to the result of the numerical simulation where the frequency was 
an increasing function of $v$.  
The reason of this discrepancy is not clear at the moment.  
Presumably the above model will be too much simplified to discuss such 
detailed behaviors as delicate changes in inner degrees of freedom of the 
soliton.  

We have found in the numerical simulation that the amplitude mode is excited 
with the acceleration of the soliton.~\cite{amp}  
When the soliton moves with a constant velocity, the soliton width 
oscillates with an almost constant frequency and amplitude.  
In reality, however, there will be the damping of the amplitude mode due to 
the interactions with other modes which are not included in the Lagrangian 
eq. (\ref{eq:L}), and the amplitude of the width oscillation will damp and 
then the width settles down to a certain value, which depends on the 
velocity.  
The effect of the damping could be introduced by adding a phenomenological 
friction term in the equation of motion.  
After the elapse of sufficient time, the time derivative of the width and its 
second derivative can be both treated as zero.  
The equation of motion can be reduced to
\begin{equation}
\henb{\xi} \left( \frac{M_{\rm s}(\xi)}{2} v^2 -V(\xi) \right) = 0,
\end{equation}
which is the equation to determine the width of the soliton moving with the 
velocity $v$.  
The soliton width is determined by an extreme of the function inside the 
parentheses which we call the effective action.  
The effective action eq. (\ref{eq:A_sol}) can be obtained by inverting the 
sign and introducing the relative width $r$ in the function.  

The effective action for the acoustic polaron can be derived 
by a similar way as above.

\def\PR{Phys. Rev. }
\def\PRL{Phys. Rev. Lett. }
\def\SSC{Solid State Commun. }
\def\JPS{J. Phys. Soc. Jpn. }
\def\SM{Synthetic Metals }

\newpage
\noindent
{\bf Figure Captions}
\begin{description}
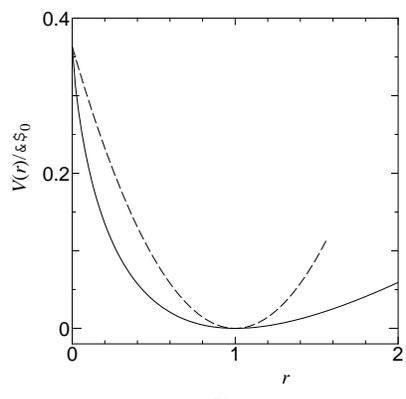
\item[Fig. 1 ] The potential energy $V(r)$ for the case of the  soliton 
calculated from the SSH model (solid line).  
$V(r)$ is equivalent to the creation energy of a static soliton as a function 
of its width.  
The energy is scaled by $\Delta_0$ (half of the Peierls gap)  
and the width by its ground state value 
$\xi_0$.  
The deviation in energy from the ground state energy is plotted.
The broken line indicates the approximation employed by Bishop 
{\it et. al}.~\cite{Bishop}  

\item[Fig. 2 ] The action (upper part) and its derivative by $r$ (lower part) 
for $\beta_v < \beta_{\rm c}$ (a), $\beta_v = \beta_{\rm c}$ (b), and 
$\beta_v > \beta_{\rm c}$ (c).  
In each case the potential energy (broken line) and the kinetic energy 
(dash-dotted line) are shown in upper part.

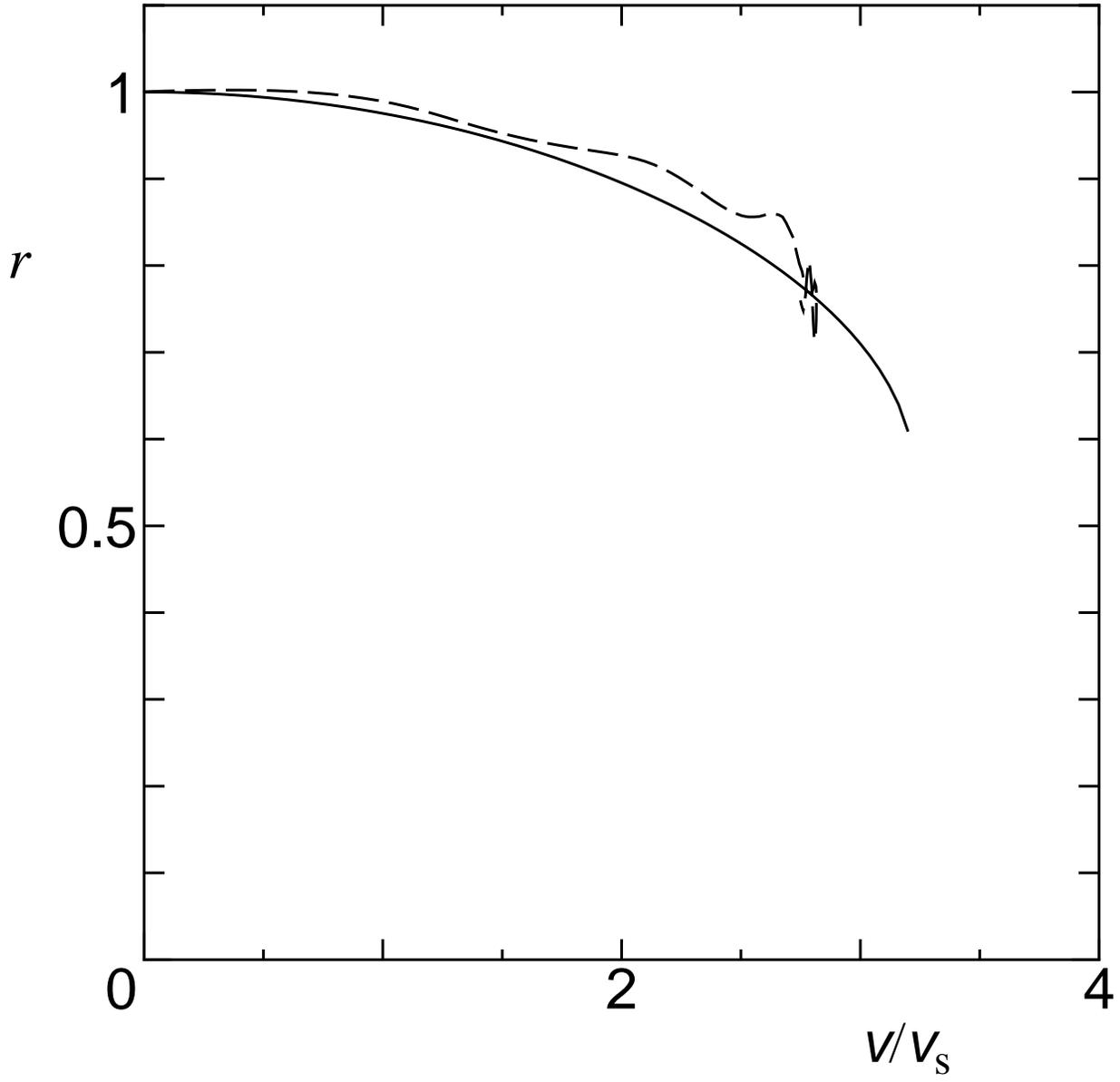
\item[Fig. 3 ] The velocity dependence of the soliton width determined from 
the effective action (solid line) and from the numerical simulations (broken 
line).  
The width $\xi$ is scaled by $\xi_0$ and the velocity by the sound velocity.

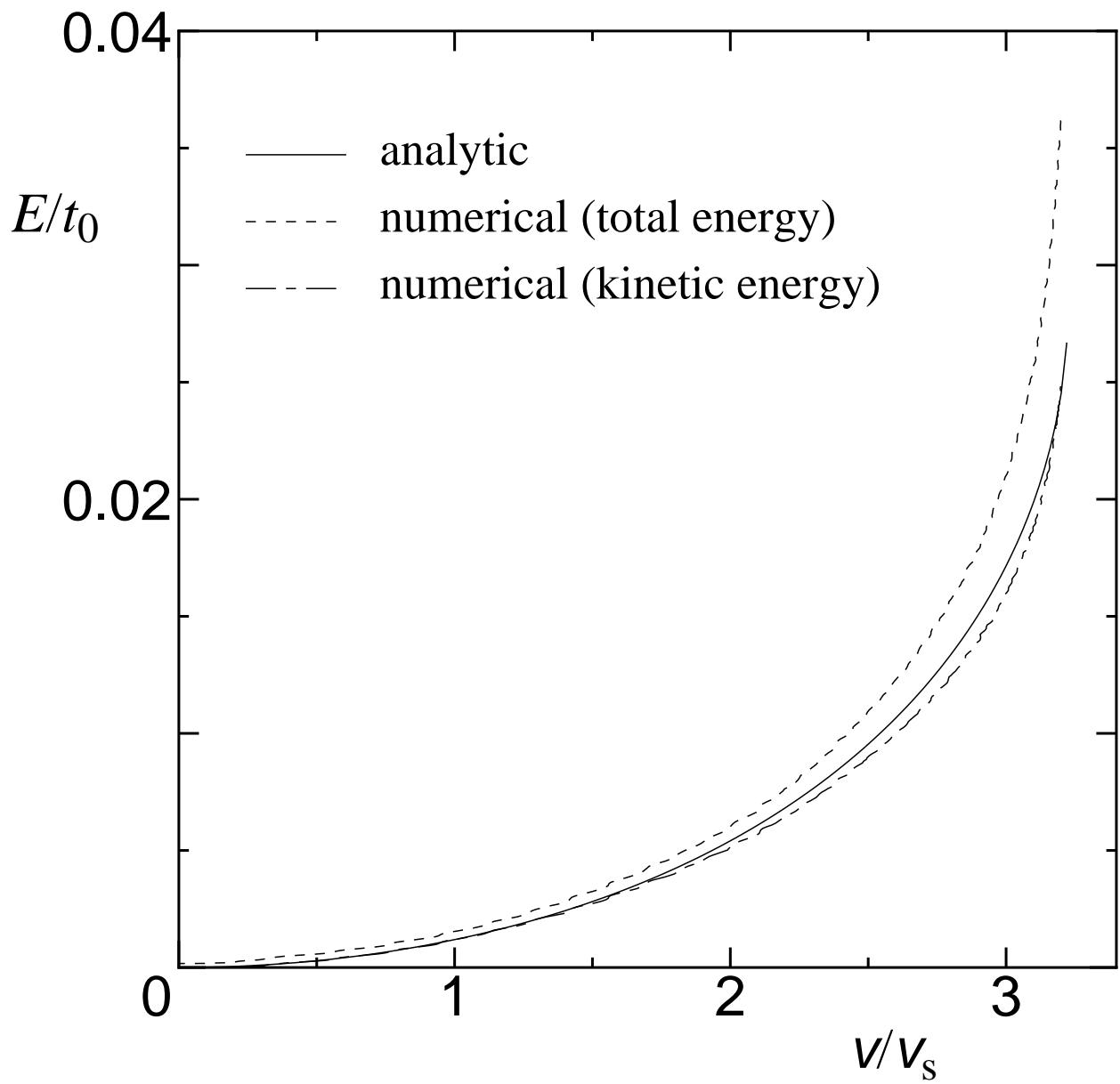
\item[Fig. 4 ] The velocity dependence of the soliton energy determined from 
the effective action (solid line) and from the numerical simulation (broken 
line).

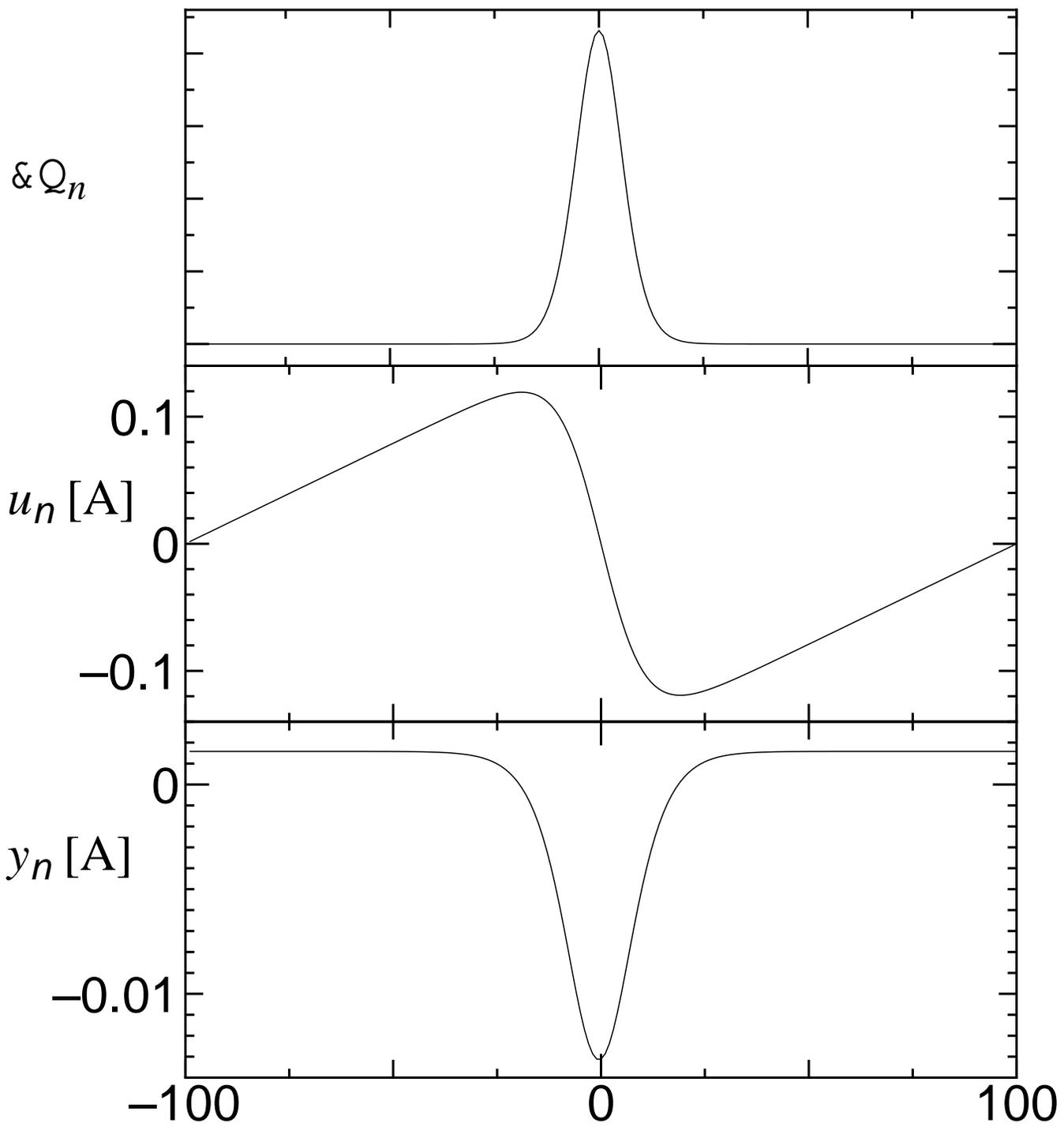
\item[Fig. 5 ] Structure of the acoustic polaron for the case of the finite 
system size with periodic boundary condition.  
The charge density (top), the lattice displacement (middle) and the bond 
configuration 
(bottom) are shown as functions of the site number.

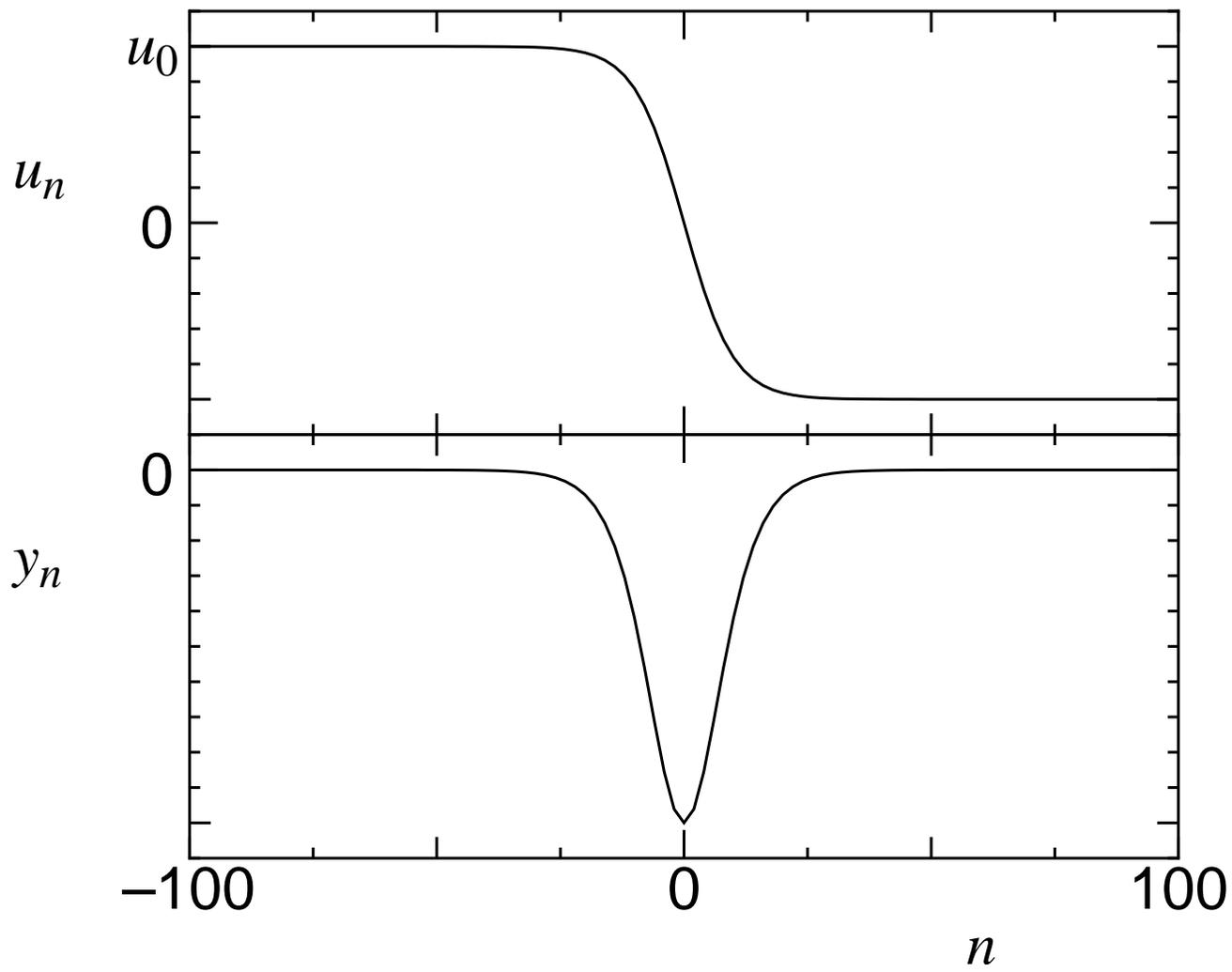
\item[Fig. 6 ] Structure of the acoustic polaron for the case of the infinite 
system size.  
The lattice displacement (top) and the bond configuration (bottom) are shown.  
The ordinates are in arbitrary unit.

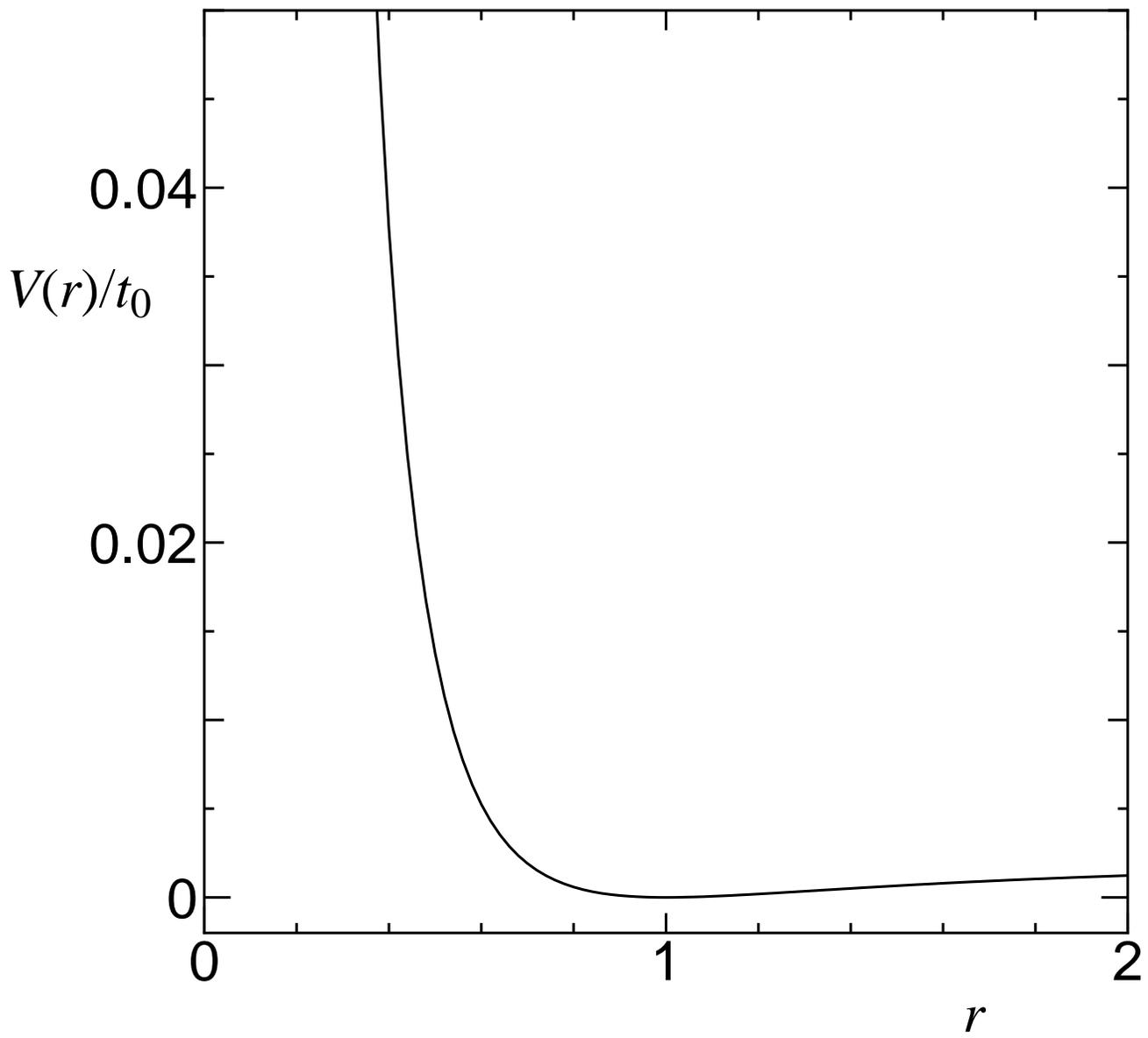
\item[Fig. 7 ] The potential energy for the case of the acoustic polaron.

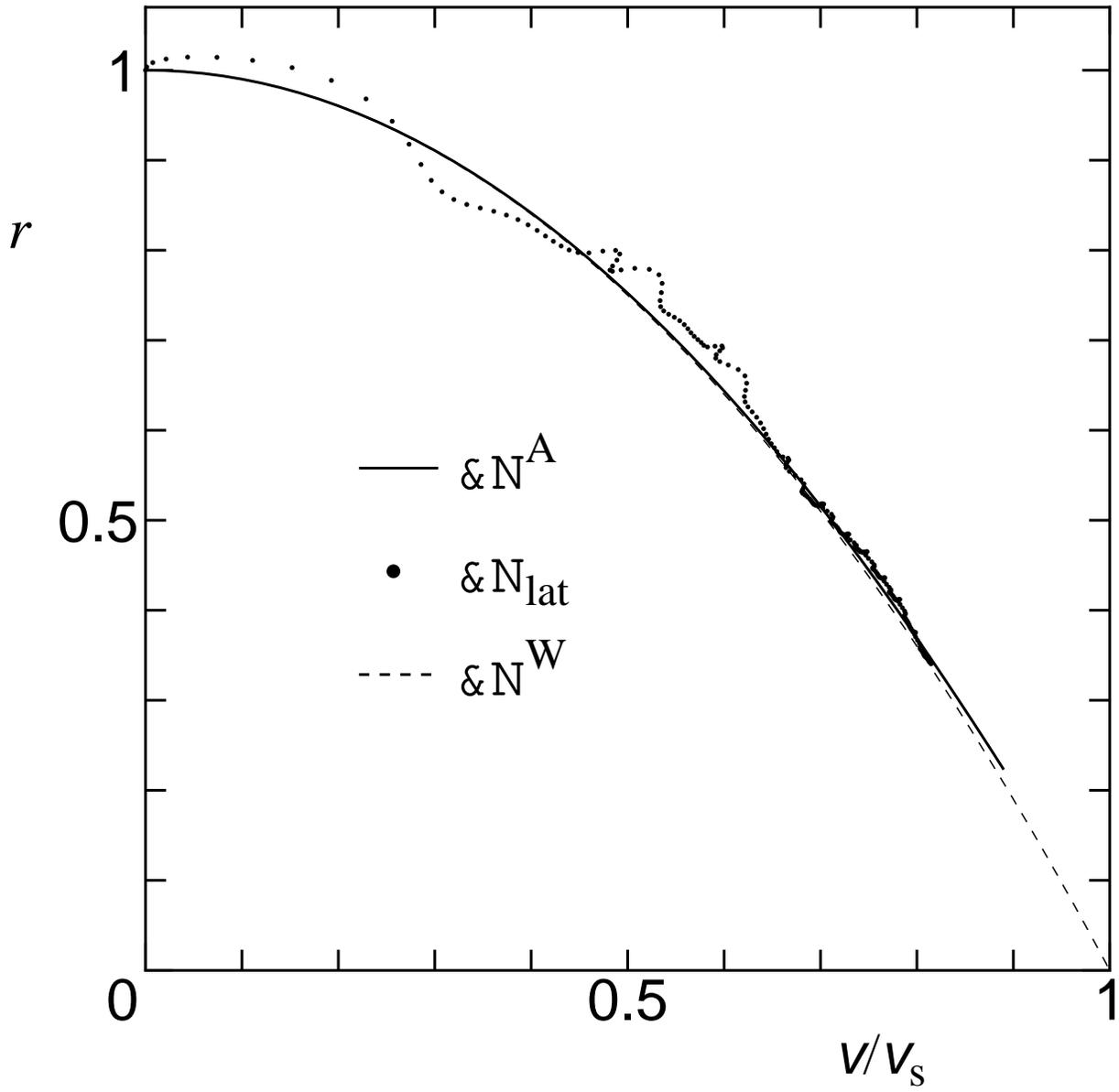
\item[Fig. 8 ] The velocity dependence of the width of the acoustic polaron 
determined from the effective action $\xi^{\rm A}$ (solid line) and from the 
numerical simulation of the polaron motion $\xi_{\rm lat}$ (solid dots).  
The broken line indicates Wilson's result $\xi^{\rm W}$.

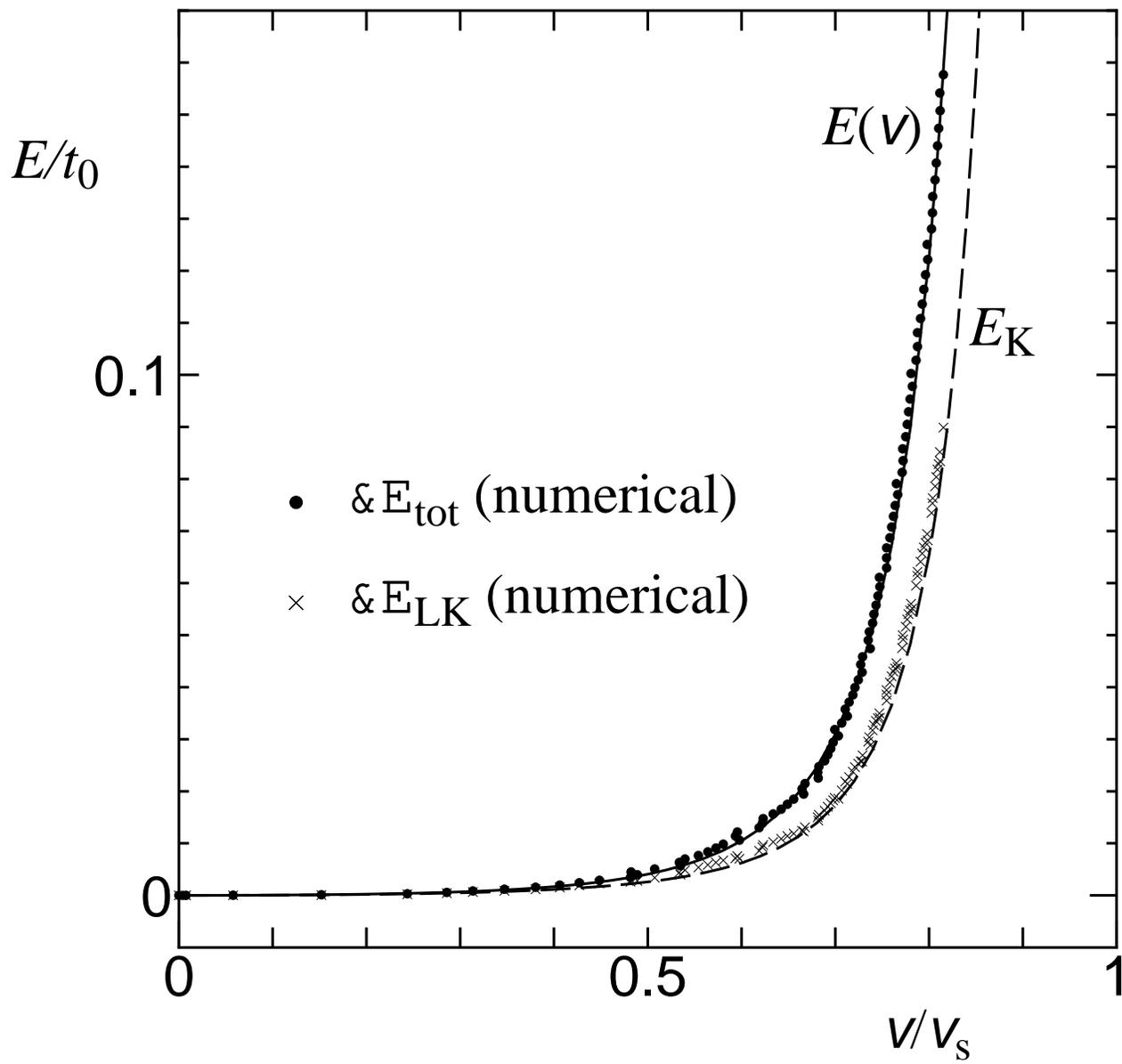
\item[Fig. 9 ] The velocity dependence of the energy of the acoustic polaron 
(solid line) and its kinetic energy part (broken line) determined from the 
effective action.  
Those determined by the numerical simulation are indicated by dots 
and crosses, respectively.
\end{description}

\begin{thebibliography}{99}
\bibitem{footnote1}
This expression differs from what is obtained by expanding $V(r)$ 
around $r=1$ up to the second order of $(r-1)$.
\bibitem{footnote3}
Since the potential energy in the effective action is calculated 
numerically in the case of the acoustic polaron, it may not be 
completely appropriate to call the result evaluated from the 
effective action ^^ ^^ analytic result''.  
However we use terminology ^^ ^^ analytic'' in order to emphasize that 
the numerical simulation is not necessary to analyze the structure
of the moving nonlinear excitations.
\end{thebibliography}
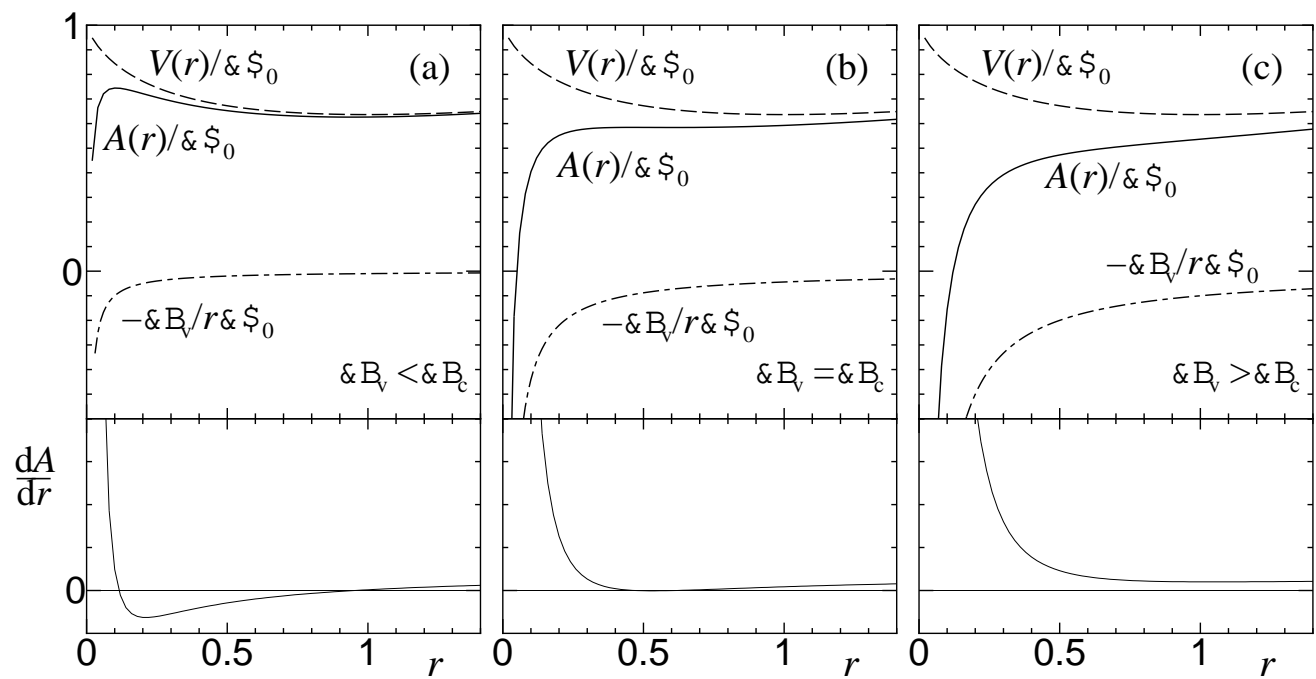
\end{document}